\documentclass[preprint, authoryear,times]{elsarticle}

\usepackage{graphicx}

\usepackage{amssymb}
\usepackage{lineno}
\usepackage{lscape}

\journal{The Journal of Systems and Software}

\begin{document}

\begin{frontmatter}

\title{Towards a Quantitative Measurement of Agility: A Validation Study on an Agile Maturity Model}

\author[torkar-new]{Lucas Gren\corref{cor1}}
\ead{lucas.gren@cse.gu.se}
\cortext[cor1]{Corresponding author. Tel.: +46 739 882 010}

%\author[torkar-new]{Richard Berntsson Svensson}
%\ead{richard@cse.gu.se}

\author[torkar-new,torkar]{Richard Torkar}
\ead{richard.torkar@cse.gu.se}

\author[torkar,torkar-new]{Robert Feldt}
\ead{robert.feldt@bth.se}

\address[torkar-new]{Chalmers University of Technology and the University of Gothenburg, SE-412 96 Gothenburg, Sweden}
\address[torkar]{Blekinge Institute of Technology, SE-371 79 Karlskrona, Sweden}

\begin{abstract}
Agile Development has now become a well-known approach to collaboration in professional work life. Both researchers and practitioners want validated tools to measure agility. This study sets out to validate an agile maturity measurement model with statistical tests and empirical data. First, a pretest was conducted as a case study including a survey and focus group. Second, the main study was conducted with 45 employees from two SAP customers in the US. We used internal consistency (by a Cronbach's alpha) as the main measure for reliability and analyzed construct validity by exploratory principal factor analysis (or PFA). The results suggest a new categorization of a subset of items existing in the tool and provides empirical support for these new groups of factors. Since there are few validated tools to measure agile maturity, researchers cannot correlate quantitative agile maturity measurements to other variables in Software Engineering research and be confident that the results are correct. Practitioners cannot either use these tools to guide their journey towards agility. Based on the results we suggest a new partly validated set of items from the tested tool. However, we argue that maturity models do not measure agility in a way that is desired. 

\end{abstract}


\end{frontmatter}

\section{Introduction}
The study of agile development and management practices is a relatively new field of research. The term itself, ``agile development'', was first coined in the area of software development but similar concepts preceded it in the literature on manufacturing. Today it has become a general project management concept\slash tool, and the word ``agile'' is frequently used in the general business and project management literature, e.g.\ \citet{miles, poolton, vinodh}.

Agile methods in software engineering evolved during the 1990s and in 2001 it became a recognized concept due to ``The manifesto for agile software development'' written by a group of software developers \citep{fowler2001}. According to \citet{agilecobb} the background to the agile ideas was that projects in crisis sometimes took on more flexible ways of thinking and working and then were more successful. This style was named ``agile'', which literally means to be able to move quickly and easily \citep{fowler2001}, and emerged in reaction to more traditional project management methods were detailed planning typically precedes any implementation work.

During the 1990s the traditional way of doing procurement, elicitation of requirements, contract negotiations and then production and, finally, delivery (e.g.\ what is often termed the waterfall model in software development literature), sometimes helped create computer and software systems that were obsolete before they were delivered. To try to solve these challenges the agile community thus defined a set of principles that they summarized in The Agile Manifesto \citep{fowler2001}:

\begin{itemize}
\item Individuals and interactions over processes and tools.
\item Working software over comprehensive documentation.
\item Customer collaboration over contract negotiation.
\item Responding to change over following a plan.
\end{itemize}

\citet{laanti} claim that scientific and quantitative studies on agile methods were still rare in 2011, while requesting such studies since they can give more general advice about the practices involved. Overall, if an organization wants to transition to more agile ways of working, regardless of whether they are a software organization or not, the decision-makers will benefit from measuring agility both before, during, and after such a transition. The question is if this is possible since agility is a cultural change (described in the agile manifesto above) as well as a smorgasbord of practices to support them \citep{williams,doingtobeing,scrumorbeing}. 

There is a diversity of agile measurement tools out there, both scientific and commercial but almost none of them has been statistically validated. In order to measure measure agility and trust in the given results\slash output, both researchers and practitioners need validated tools to guide their process. The problem is what to focus on and on what level, since the agile approach is on a diversity of levels in the organization. This empirical study will evaluate one of the agility maturity models found in research through a statistical validation process. This tool focuses a bit more on behavior and not only lists a set of practices for the research subjects to tick yes or no regarding if they are implemented or not. We also connect a Likert scale to the evaluation in order to capture more variance in connection to each item. Section~\ref{sec:background} will outline existing agile measurement tools found in the literature, Section~\ref{sec:method} will present how our main statistical investigation was conducted, but also describe a pretest conducted before the main study including its findings under Subsection~\ref{sec:prestudy}, Section~\ref{sec:results} will present main study findings, Section~\ref{sec:discussion} will analyze and discuss these overall results, and, finally, Section~\ref{sec:future} will present conclusions and suggest future work.

This study aims to contribute with the following: 
\begin{enumerate}
\item A test to evaluate if the Agile Adoption Framework can be used to measure current agility (instead of agile potential).
\item If practitioners think such an evaluation is relevant through a case study pretest.
\item Expand the Agile Adoption Framework to include a Likert scale evaluation survey filled out by all the team members and not just by the assessor\slash researcher and connect a confidence interval to the item results.
\item Partly validate the Agile Adoption Framework with statistical tests.
\item Suggest changes Agile Adoption Framework and\slash or highlight the issues connected to agility measurement.
\end{enumerate}

\section{Related Work}\label{sec:background}
Some researchers suggest qualitative approaches like interviewing as a method for assessing agility in teams \citep{boehm20, sidky, pikka}. \citet{hoda} even suggest the use of grounded theory which is an even more iterative and domain specific analysis method \citep{GlaserS67}. Interviewing is a good way to deal with interviewee misinterpretations and other related biases. The work proposed by \citet{lee} compares a few agility dimensions with performance and draw conclusions about the complexity of if agile methods increase performance or not, which they do.

\citet{datta} describes an Agility Measurement Index as an indicator for determining which method of Waterfall, Unified Software Development Process (UP), or eXtreme Programming (XP) should be used. Where Waterfall is plan-driven and XP is an agile method, UP is considered to have elements of both and is a more general framework that can be adapted to specific needs but that is often used as a kind of middle ground between the other two. The author suggests that the five dimensions: Duration, Risk, Novelty, Effort, and Interaction should be taken into account when selecting development method. Their method is, however, a company-specific assessment, which makes comparisons between different organizations cumbersome.

A process control method often used within IT is the CMMI (Capability Maturity Model Integration). This method also divides the organization into different maturity levels and is essentially a set of requirements for engineering processes, particularly those involved in product development. Just like stage-gate project management these older methods often co-exist with agile methods when implemented \citep{turner20022}, i.e.\ CMMI is constructed to measure a more traditional project management approach.

To be able to compare and guide organization in their agile implementations a diversity of agile maturity models have been suggested, as mentioned in the introduction. \citet{lepp} presents a nice overview of these agile maturity tools selected with the following criteria: ``Domain'' (the domains the models are targeted to), ``Purpose'' (the purposes the models have been developed for), ``Conceptual and Theoretical Bases'' (the conceptual and theoretical backgrounds upon which the models have been built), ``Approaches and Principles'' (the approaches and principles used to construct the models), ``Structure'' (the architectures of the models), and ``Use and Validation'' (extent of deployment and validation). Based on these criteria eight tools were selected: The Agile Maturity Model \citep{ambler}, A Road Map for Implementing eXtreme Programming \citep{lui}, Toward Maturity Model for Extreme Programming \citep{toward}, The Agile Maturity Map \citep{packlick}, Agile Maturity Model \citep{patel}, Agile Maturity Model \citep{pettit}, A Framework to Support the Evaluation, Adoption and Improvement of Agile Methods in Practice \citep{qumer}, and The Agile Adoption Framework \citep{sidky}. According to \citet{lepp} some of them are merely based on conceptual studies, others are developed only in one organization, a third group has gathered more experience from organizations, and some are discussed with practitioners. However, as also \citet{lepp} concludes, none of them are validated. He also states that higher maturity levels could partially be assessed by more lightweight methods. 

In this study we selected to focus on the Sidky's Agile Adoption Framework, and in order to keep the number of items as low as possible, we selected only level one of this tool. We should also mention that there is a set of commercial tools available, however, their scientific foundation is hard to assess. 

We would like to highlight the difficulty of measuring something that is an ambiguous construct, such as agility. Maturity is of course even harder to assess in connection to agility since maturing with a unspecific concept is even harder. However, there are some behaviors connect to ``being agile'' in software development and behavior connected to this way of working, which is our definition of agile maturity in this case. We do not aim to find a way to quantitatively measure agility in this study (and we neglect the agile practices effectiveness as well), but instead to test one of the existing tools and try to understand how to proceed in measuring\slash dealing agility transformations in organizations.

\subsection{Sidky's Agile Adoption Framework}\label{sub:agile_adoption_framework}
In order to determine which agile methods an organization is ready to use, \citet{sidkyphd} suggests a method called the Agile Adoption Framework. He motivates its use by arguing that even though there are many success stories in agile development, they are not really generalizable, i.e.\ it is unclear how the case by case descriptions can be used to judge agility readiness for a company which has some, but not all, aspects in common with reported cases. \citeauthor{sidkyphd} also criticizes more general frameworks, since they address agility in its generic form and not the actual practices.

Sidky's approach is based on a tool that has two parts. The first part is called the Agile Measurement Index (the same name as \citet{datta} uses, but a different tool) and is:

\begin{itemize}
\item A tool for measuring and assessing the agile potential of an organization independent of any particular agile method (based on behavior connected to practices that fit into the agile manifesto).
\item A scale for identifying the agile target level will ultimately aim to achieve.
\item Helpful when organizing and grouping the agile practices in a structured manner based on essential agile qualities and business values.
\item Able to provide a hierarchy of measurable indicators used to determine the agility of an organization.
\end{itemize}

The second part is the use of the Agile Measurement Index through a four-stage process that will assess, firstly, if there are discontinuing factors (factors that result in a no-go decision, i.e.\ deal-breakers), secondly, an assessment at project level, thirdly, an organizational readiness assessment, and lastly, a reconciliation phase. The main purpose of the tool as a whole is to guide organizations on which agile practices to adopt. We only use the first part from this framework later on, since we only want to measure agile practices.

The Agile Adoption Framework is divided into agile levels, principles, practices and concepts, and indicators. The concept of an agile level collects a set of practices that are related and indicates the degree to which a core principle of agility is implemented. An agile principle is a set of guidelines that need to be employed to ensure that the development process is agile; the principles used are derived from the basic and common concepts of all agile methods. The agile practices and concepts are tangible activities that can be used to address a certain principle. (Table~\ref{fig:alp} shows the agile principles and their practices on the different levels.)

The indicators are items used to assess certain characteristics (see Table~\ref{fig:ratcp}) and are collected through interviews with representatives from different parts of the organization. In order to assess what agile level is suitable for a group to implement, the indicators are used and summarized as a percentage to what degree they are achieved. The same type of assessment and summary is performed also at the organizational level. The last phase (reconciliation) is when these two assessments, the group and organizational levels, are mapped together to decide what agile practices to implement in projects in relation to the organization around them \citep{sidky}.

\begin{table*}
\caption{Agile Levels, Principles, and Practices \citep{sidkyphd}}
\label{fig:alp}
\centerline{\includegraphics[scale=0.6]{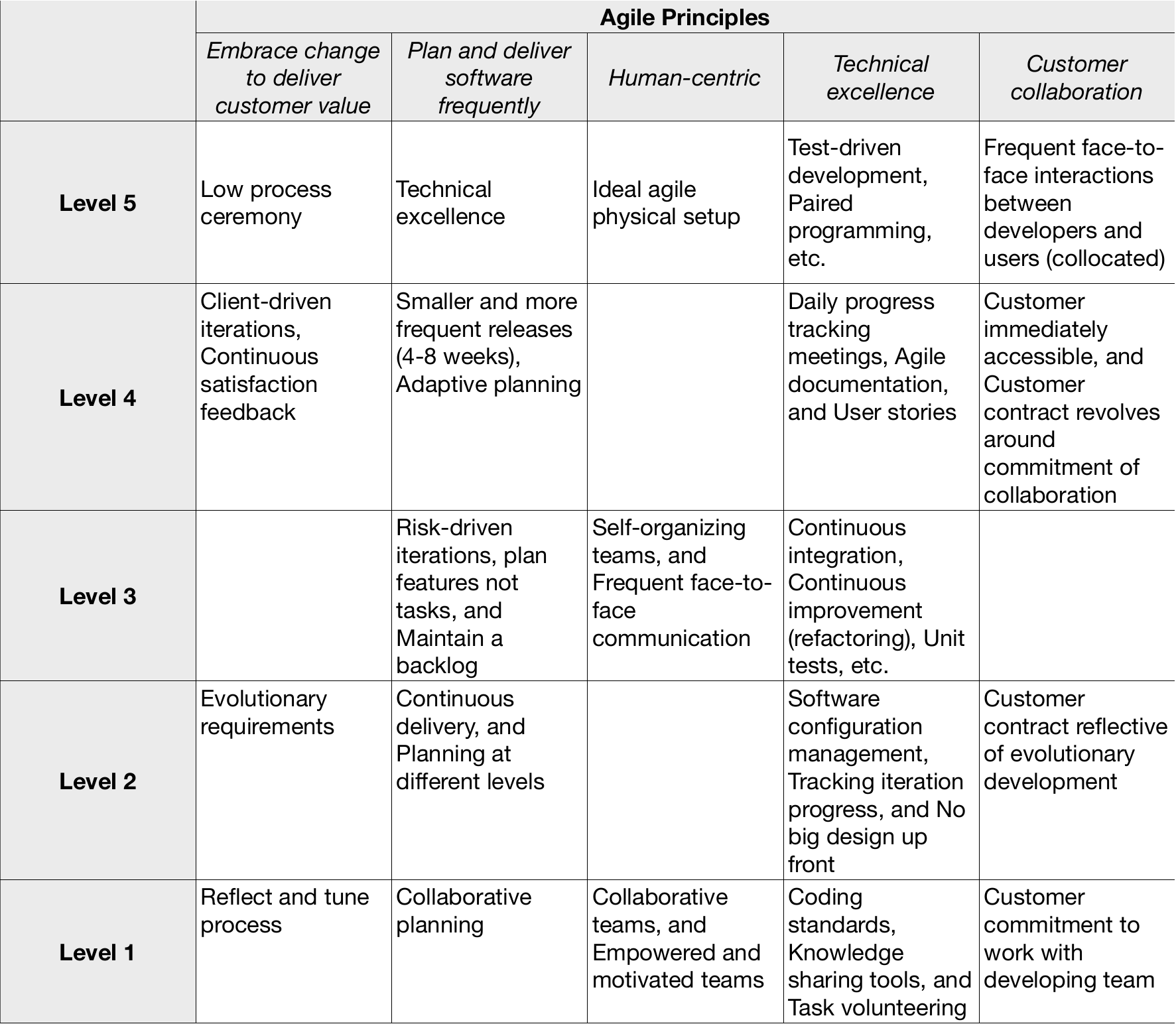}}
\end{table*}

Sidky defines ``how agile'' a company is by the amount of agile practices they use. This makes a measurement tool possible and straightforward, and means that an organization that uses ten agile practices is considered to be more agile than one that uses three. The indicators are then connected to these practices and divided into respondent groups such as developers, managers and assessors, but the assessors do all the evaluations on a Likert scale from 1 (Strongly disagree) to 5 (Strongly agree) based on interviews. These responses are grouped in order to answer the characteristic to be assessed (Table~\ref{fig:ratcp} shows all the indicators used to assess the agile practice ``Collaborative Planning'' as seen in Table~\ref{fig:alp} in the last row, third column from left). 

\begin{landscape}
\begin{table}
\caption{Readiness Assessment Table for Collaborative Planning \citep{sidkyphd}}
\label{fig:ratcp}
\centerline{\includegraphics[width=180mm]{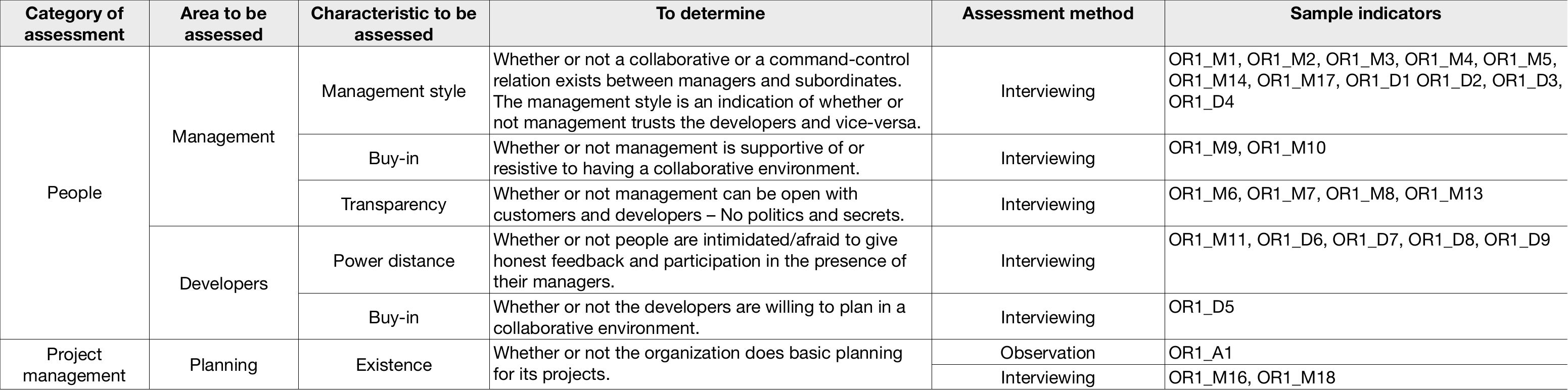}}
\end{table}
\end{landscape}

Table~\ref{fig:ratcp} also states what data collection technique should be used to assess each indicator.

Sidky sorts all practices in different agile levels depending on how ``advanced'' they are. We think this division of practices is arbitrary but for simplicity we have chosen to evaluate our method at a level corresponding to Level 1 to keep the number if items to a minimal. Table~\ref{fig:sidkyassess} shows all the agile practices assessed at Level 1. Each characteristic is evaluated through a combination of indicators taken from both developer and manager interviews. Below Table~\ref{fig:sidkyassess} you will also find a a description of what the agile characteristics set out to determine.

\begin{table*}
\centering
\caption{Descriptions of What the Different Characteristics Set Out to Determine \citep{sidkyphd}}
\includegraphics[width=\textwidth]{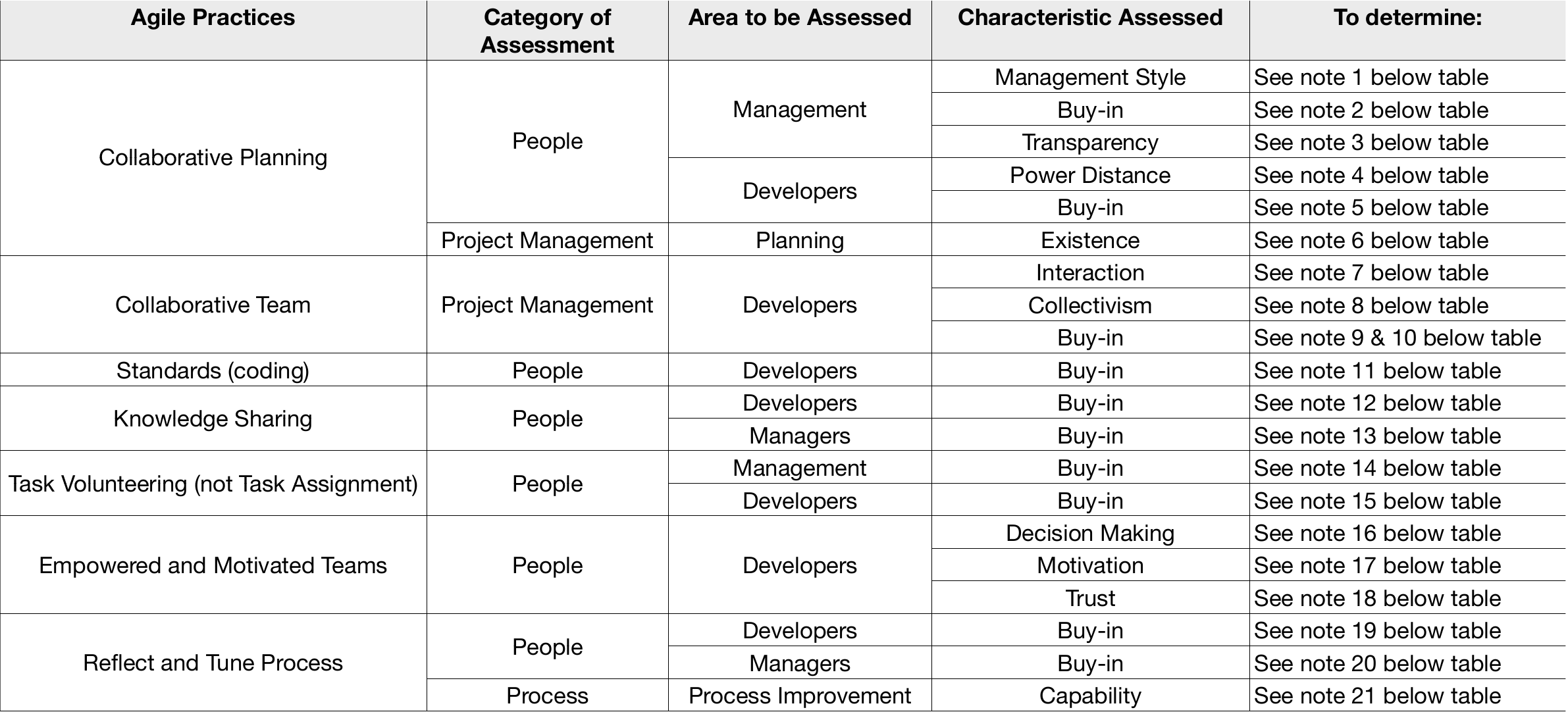}\label{fig:sidkyassess}
\begin{enumerate}
\scriptsize{
\item Whether or not a collaborative or a command-control relation exists between managers and subordinates. The management style is an indication of whether or not management trusts the developers and vice versa.
\item Whether or not management is supportive of or resistive to having a collaborative environment.
\item Whether or not management can be open with customers and developers, i.e., no politics and secrets.
\item Whether or not people are intimidated\slash afraid to give honest feedback and participation in the presence of their managers.
\item Whether or not the developers are willing to plan in a collaborative environment.
\item Whether or not the organization does basic planning for its projects.
\item Whether or not any levels of interaction exist between people thus laying a foundation for more team work.
\item Whether or not people believe in group work and helping others or are just concerned about themselves.
\item Whether or not people are willing to work in teams.
\item Whether or not people recognize that their input is valuable in group work.
\item Whether or not the developers see the benefit and are willing to apply coding standards.
\item Whether or not developers believe in and can see the benefits of having project information communicated to the whole team.
\item Whether or not managers believe in and can see the benefits of having project information communicated to the whole team.
\item Whether or not management will be willing to buy into and can see benefits from employees volunteering for tasks instead of being assigned.
\item Whether or not developers are willing to see the benefits from volunteering for tasks.
\item Whether or not management empowers teams with decision making authority.
\item Whether or not people are treated in a way that motivates them.
\item Whether or not managers trust and believe in the technical team in order to truly empower them.
\item Whether or not developers are willing to commit to reflecting about and tuning the process after each iteration or release.
\item Whether or not management is willing to commit to reflecting about and tuning the process after each iteration or release.
\item Whether or not the organization can handle process change in the middle of the project.}
\end{enumerate}
\end{table*}

The evaluation of the results of this first part of the tool is then done in four steps. The first step is to compute a weight for each indicator. If no indicators are believed to be more important, the weight of 1 is divided by the number of indicators. If the indicators are weighted differently they must all sum to 1. The second step is to compute weighted intervals. These are done by taking the answer of each indicator and calculate a pessimistic and optimistic result. The Likert scale is then divided into a percentage according to Table~\ref{fig:normalpt}.

\begin{table*}
\caption{Optimistic and Pessimistic Percentage Table for each Rating \citep{sidkyphd}}
\label{fig:normalpt}
\centerline{\includegraphics[width=100mm]{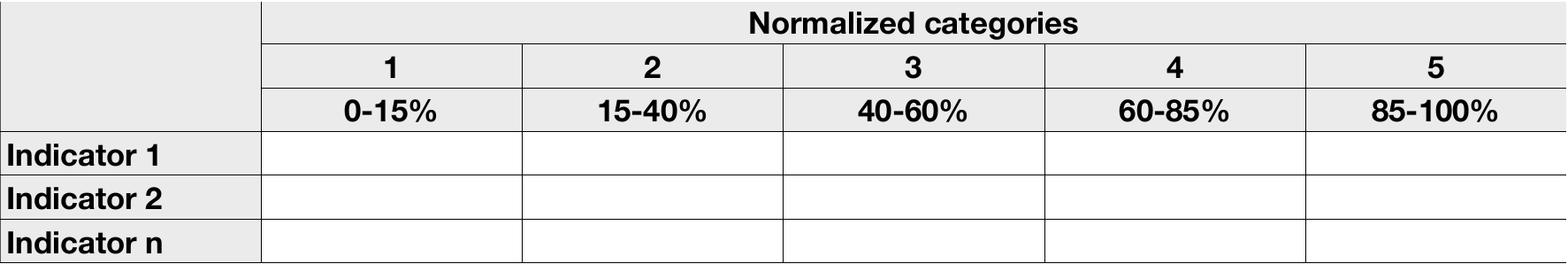}}
\end{table*}

The pessimistic (lower) result is the sum of all the weighted low-end results of the normalized categories shown in Table~\ref{fig:normalpt}. For example if the score is 4 the low-end result is 60\% and the high-end result is 85\% for the same data point. The same is done for each question with the high-end results and summed into an optimistic result. These are then compared to the nominal scores in Table~\ref{fig:nominal}. If the calculated interval is outside the nominal intervals given, an average is calculated and used instead \citep{sidkyphd}.

\begin{table}
\caption{Nominal Values for Optimistic and Pessimistic Range Comparisons \citep{sidkyphd}}
\label{fig:nominal}
\centerline{\includegraphics[width=40mm]{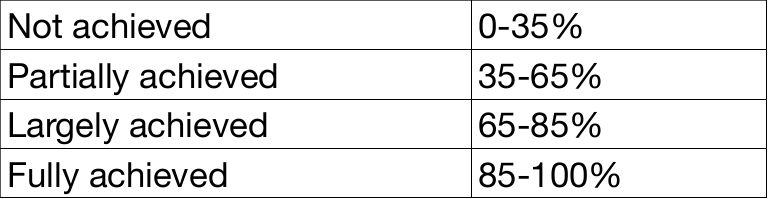}}
\end{table}
 
This tool, created by \citet{sidkyphd}, is based on interviews and assesses the level of agility an organization is prepared to implement and recommends what particular methods should be used. However, in order to make sure we collect the variance in the responses, we decided to measure teams that state they work with some agile methods already. The method of interviewing to assess agility is also time-consuming and it would be an advantage if this could be done as a survey instead. This is also, partly, necessary in order to use statistical analysis methods. \citeauthor{sidkyphd} defines agile practices and connects indicators (or items) to them according to his opinion, i.e., no statistical method was used, neither was the creation of his framework clearly based on empirical data from actual teams. He then evaluated the items by letting expert agile practitioners give their feedback on the tool. No further validation has been conducted.

This study includes two parts. First, we tested Sidky's tool on two teams at Volvo Logistics in Sweden by letting the team members fill out the survey (N=15). By doing this we received many data points for each team instead of having an assessor note one data point for each. We then fed this result back to the teams in a focus group to see if they thought it was true for their team. The second step was to use a larger sample from two other companies (N=45) to see if Sidky's (\citeyear{sidkyphd}) items group in factors in the same way as he categorizes them, i.e.\ the next step in scale construction for social interaction. If a scale is to be used a qualitative generation of items must be followed by a quantitative validation analysis \citep{giles}. In this study, we chose internal consistency as the main measure for reliability and analyzed construct validity by exploratory factor analysis.

\begin{figure}
\caption{Overview of the methodology used.}
\label{fig:method}
\centerline{\includegraphics[scale=0.3]{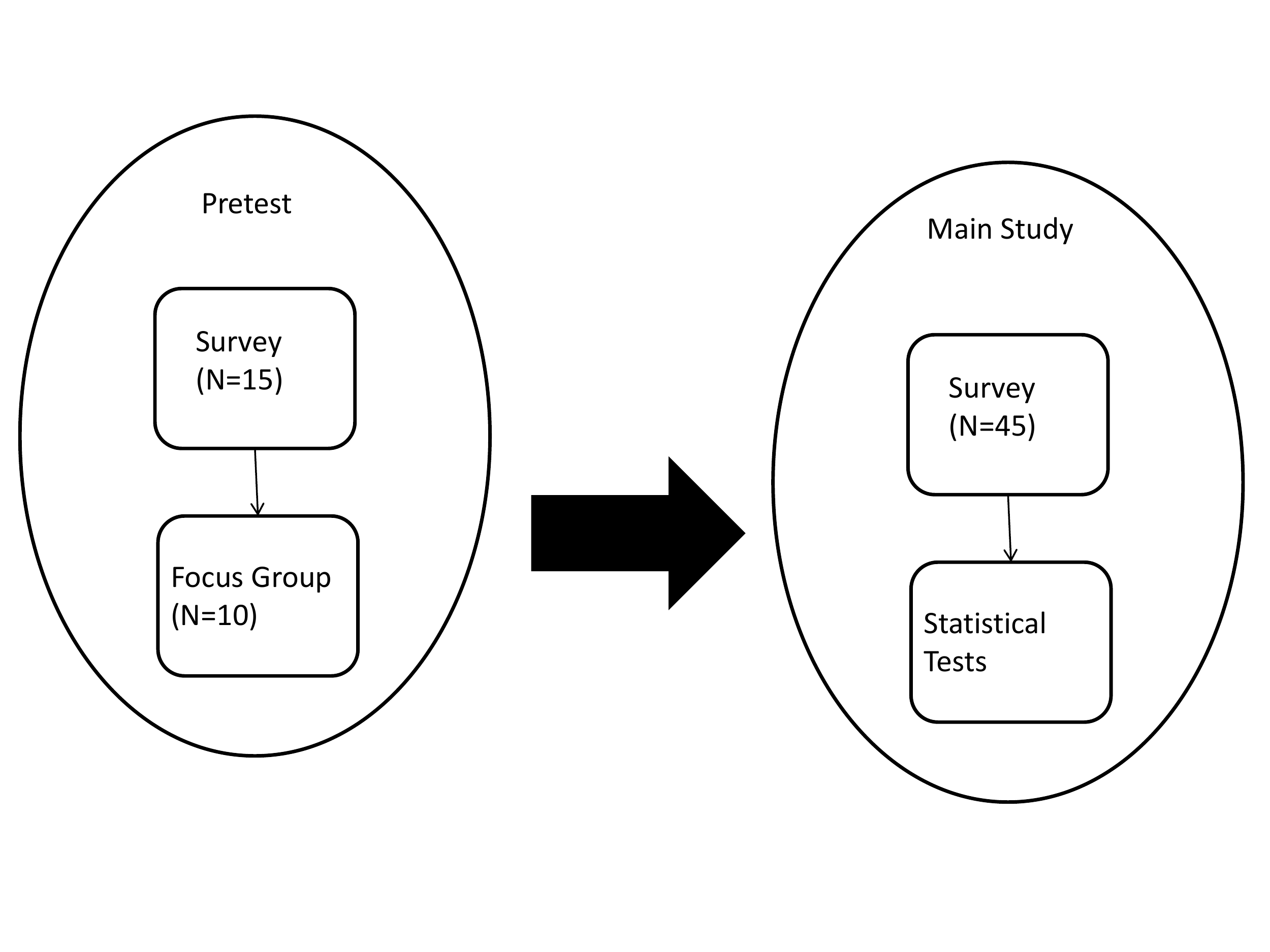}}
\end{figure}

Next we will present a pretest conducted with two teams at Volvo Logistics. This part of the study tests a survey approach to Sidky’s tool on a small sample (N=15). The purpose was to evaluate the results with the teams afterward in order to assess the appropriateness of using the tool in this manner. After this assessment we present the main methodology of the study in Section~\ref{sec:method}. We then proceed and use the tool on a large sample (N=45) and conduct statistical validation tests, which is in focus for the rest of this paper. Figure~\ref{fig:method} shows the methodology used throughout the entire paper.

\subsection{Pretest}\label{sec:prestudy}
Since the pretest aims to analyze the use of a survey tool by conducting a focus group, it comprises of two research methodologies: ($i$) A descriptive survey with the purpose of gathering quantitative data and, ($ii$) an exploratory case study with the purpose of gathering qualitative data. We ultimately believe that by using these two methods we will be able to indicate if we can collect quantitative data from the team members using the Agile Adoption Framework.

\paragraph{Pretest Case and Subjects Selection}\label{sub:case_and_subject_selection}
The teams used in this pretest, were two teams with the same manager (Scrum Master) at Volvo Logistics\footnote{http://www.volvologistics.com} in Sweden. Volvo Logistics is a part of the Volvo Group which provides world-wide supply chain expertise to a set of automotive companies. The IT part is, of course, essential for the company to function. Many organizations, independent of field, need an efficient IT department to provide good solutions for the whole organization. The organization decided to work with agile methods and were conducting a pilot study in order to later spread the methods to other departments of the organization. 

The specific teams' task was to develop a part of an enterprise software system for supply chain management. During the process they worked with agile methods, and specifically Scrum. The reason why the sample is from software engineering is that they have the most experience with agile methods and were easier to find. The project was divided into two teams with the same manager (Scrum Master) consisting of a mixture of business- and programming-focused employees. This was done in order to assert the business effects of the project and create a method that more people could use within the organization. This meant, also, that many of the team members had managerial tasks during the project. Since there were unclear lines drawn between the teams and they had the same manager (Scrum Master), we chose to analyze the data collectively for both teams.

\paragraph{Pretest Data Collection Procedures}\label{sub:data_collection_procedures}
Data was collected via a paper survey with items connected to agile principles for Level 1 of Sidky's (\citeyear{sidkyphd}) tool (see Table~\ref{fig:alp}). As this table shows, Level 1 is a set of practices that is defined as the first level of agility in the tool. 

Instead of conducting interviews with all the team members they filled out the indicators themselves in the survey on a Likert scale from one to five and the assessor observational indicators were left out. Since \citeauthor{sidkyphd}'s \citeyearpar{sidkyphd} tool has indicators on behavior connected to working with agile practices it is suitable to let the team members fill out the evaluation themselves instead of having one person do the assessment after an interview. The other studies that aim to measure agility simply state an agile principle, which forces the assessor to explain these concepts so all members know how to assess them (thus introducing the risk of biasness). This also makes it possible to statistically create a confidence interval for the result based on the $t$-distribution as descriptive statistics, since a sample of many individuals is collected instead of just one. This, also, captures the deviation from the mean and the result for an indicator can then be given with a probability as confidence interval (see next section for a more thorough explanation of the procedure). 

The survey was handed out in paper form to 23 team members in the two teams and 15 filled them out. The surveys were filled out at the workplace and were anonymous. The teams had many members with managerial tasks, which make the manager sample size $(N=7)$ almost equally large as the one for developers $(N=8$). The level of agility is, in this case, a combined level for the 15 individuals that responded to the survey. After the survey results were summarized a focus group was conducted with 10 of the individuals that had filled out the surveys. In the focus group, the participants discussed the results and gave their opinions on its relevance. These points were written down and summarized.

\paragraph{Pretest Analysis Procedures}\label{sub:analysis_procedures}
Unlike \citet{sidkyphd} all the mean values from the surveys for each individuals were calculated for each item and then, the mean value of all indicators needed for a characteristic (e.g.\ ``Collaborative Planning - Management Style'') were transformed into a percentage with a 95\% confidence interval (also reported as a percentage).

To clarify, for example if 10 people responded to all the items included in the evaluation of ``Collaborative Planning - Manager Buy-in'' a mean was calculated for each of these items. In order to then assess the whole characteristic the new mean value was calculated from all the mean values used in that characteristic. So all the mean values from Table~\ref{fig:cpms} were used to get the total mean in Table~\ref{fig:sdcpms}. The standard deviations were of course used to get the confidence interval for the new mean value. To get the table in Table~\ref{fig:result}, the lower, upper, and mean values were divided by five (the maximum score) so they could be presented as a percentage.

\begin{table}
\caption{Indicators for ``Collaborative Planning - Management Style''}
\label{fig:cpms}
\centerline{\includegraphics[scale=1]{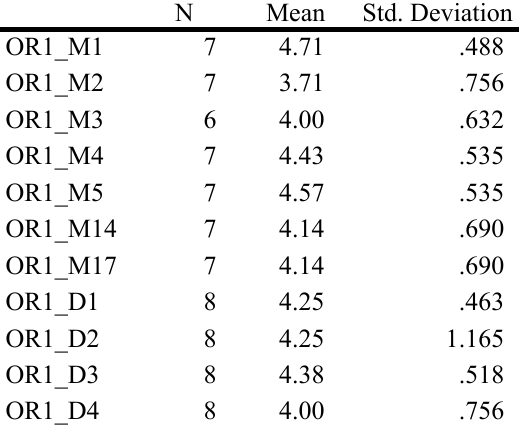}}
\end{table}
\begin{table}
\caption{Summarized Data for the characteristic ``Collaborative Planning - Management Style''; the confidence interval was calculated from a $t$-distribution with $df=7$}
\label{fig:sdcpms}
\centerline{\includegraphics[scale=1]{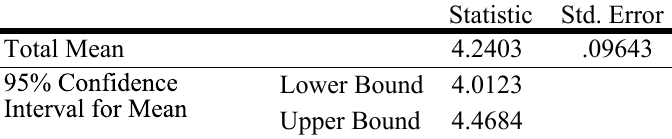}}
\end{table}

When the results were summarized, the focus group was used in order to evaluate how well the results fit reality according to the team members and the managers. This focus group was a subset of the people (10 individuals, both managers and developers) that had filled out the surveys. As mentioned before, a total of 15 individuals responded to the survey (of 23) which gives a response rate of 65\%.

\newpage
\paragraph{Pretest Results and Analysis}\label{sec:results}

\subparagraph{Summary from the surveys}
The results from the eight people replying to the survey for developers (29 items) is shown in Table~\ref{fig:datad}, and results from the seven people replying to the survey for managers (26 items) is shown in Table~\ref{fig:datam}. The total number of respondents was 15 and one manager did not reply to two items (we have not investigated the reasons for this further).

\begin{table}
\caption{Descriptive Statistics for the Survey for Developers}
\label{fig:datad}
\centerline{\includegraphics[scale=1]{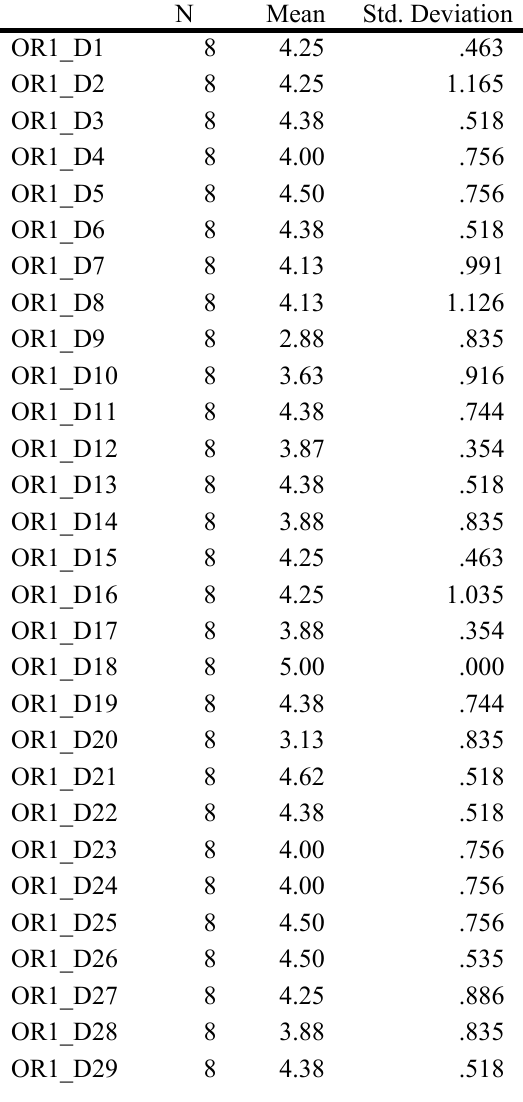}}
\end{table}

\begin{table}
\caption{Descriptive Statistics for the Survey for Managers}
\label{fig:datam}
\centerline{\includegraphics[scale=1]{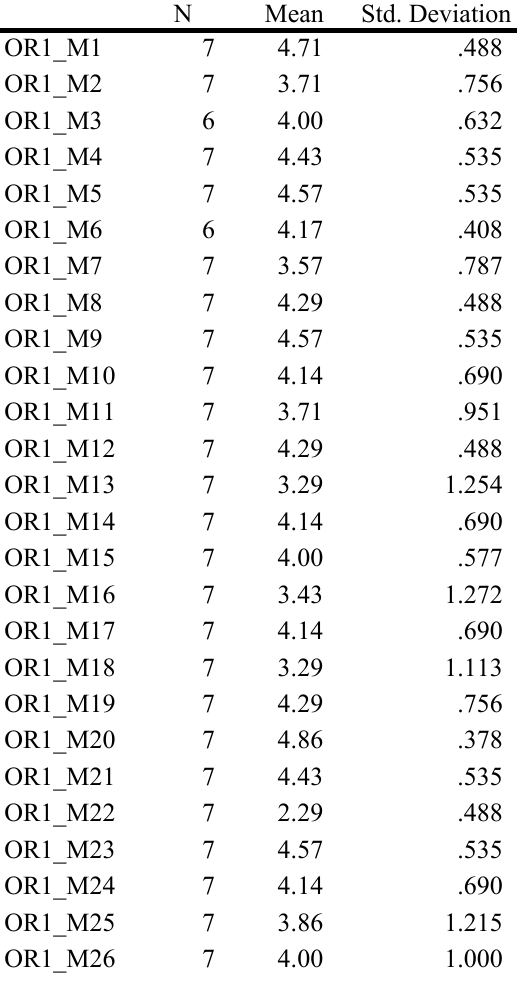}}
\end{table}

In order to get the interval to compare to nominal scores, the indicators belonging to each assessment category were calculated according to the previously described procedure, with one alteration to the tool. The alteration was based on the result of the items: OR1\_D9 and OR1\_M11 (Other peoples' titles and positions intimidate people in the organization). The results from these indicators were inverted, since the aspect of intimidation of titles must be seen as an unfortunate thing when working in agile manner. It is also stated by \citet{sidkyphd} that this item is used to determine: ``Whether or not people are intimidated\slash afraid to give honest feedback and participation in the presence of their managers'', which provides further indication that the scale should be inverted. This was also later confirmed by Sidky in email correspondence. The results of all the agile practices on Level 1 are presented in Table~\ref{fig:result}.

\begin{landscape}
\begin{table}
\caption{Results for the Studied Teams}
\label{fig:result}
\centerline{\includegraphics[scale=0.75]{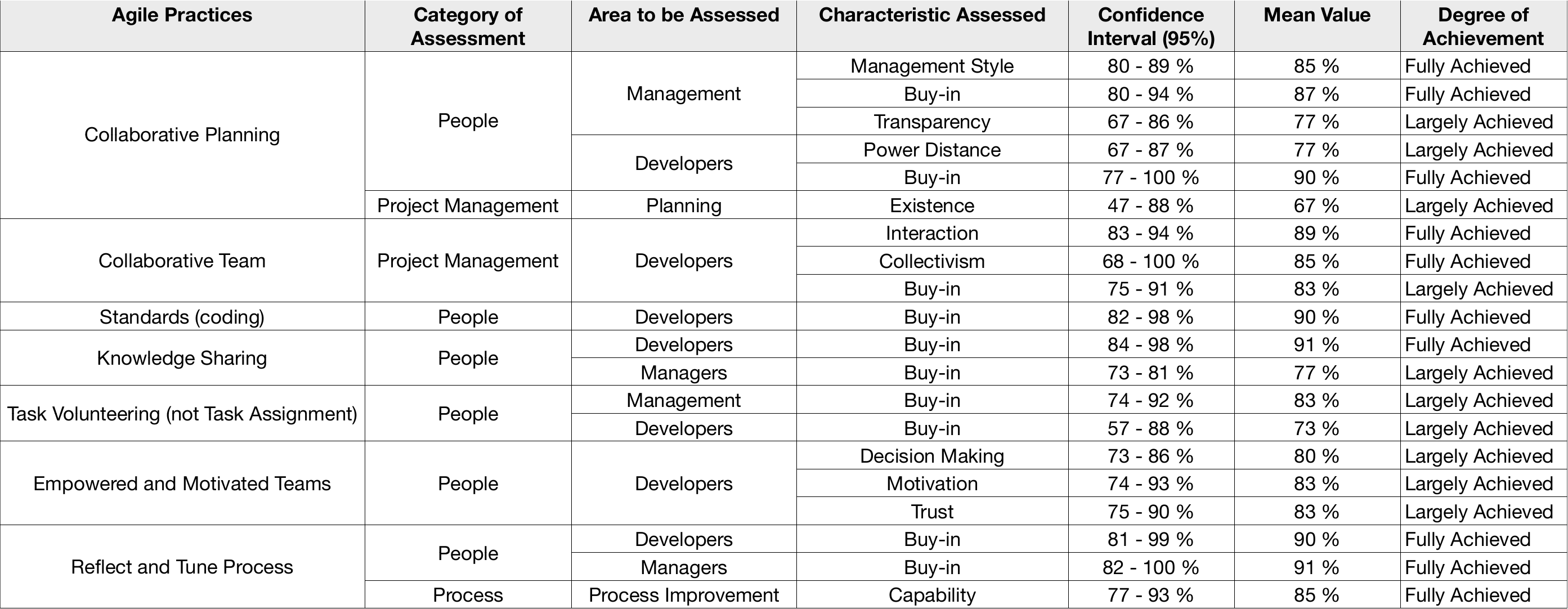} }
\end{table}
\end{landscape}

We also did a $t$-test to see if the were any differences between how managers and developers assessed the agility level. We found no such difference ($t_7=-.701, p=.495$). The reason why we did not conduct a non-parametric test was that, since the $t$-test showed no difference, neither would such a test since they are more restrictive. 

\subparagraph{Summary from the focus group}
The results were shown to the focus group and the group agreed on most results. The Scrum Master was a bit concerned that the result tended to be higher than his own expectations of the teams, but the focus group expressed that they were able to respond honestly and had done so on all items. After discussing this the Scrum Master agreed and revoked this comment. The questions about planning came up and according to \citet{sidkyphd} the items are to determine if basic planning exists. When measuring the agility of a team that tries to work agile, all members were confused if planning was good or bad. They learned to be more flexible and filled out these questions in a very different way. The focus group agreed that the questions should be altered to include ``deliverables'' instead of ``planning''. This would most likely solve the confusion regarding project planning.

Another result that was low ranked was ``task volunteering'' for the developers. The tool caught the confusion they had whether they could volunteer for tasks or not. This was because of the team consisted of both a business- and a development-focused employees, i.e., they had different roles and did not want to take tasks belonging to someone else.

As can be seen in Table~\ref{fig:result} the teams that were investigated had high results on most aspects of the surveys. This could simply be due to the fact that the teams were functioning well seen from an agile perspective. We also only used the first level of \citeauthor{sidkyphd}'s \citeyearpar{sidkyphd} tool, which could also explain the high scores. Where there were some issues, the tool caught these aspects in the variance of the result. Since this would not have shown in Sidky's tool, this motivates letting the team fill out the surveys themselves and hence collect variance in the replies and then investigate this further.

The aspects discussed in the the focus group show that \citeauthor{sidkyphd}'s \citeyearpar{sidkyphd} Agile Adoption Framework is suitable for measuring current agility in project, if the suggested alterations are made. The reason for this is that the issues discussed in the focus group and in the interview were all visible in the survey, either in the form of a low score, or with large variance associated to it.

Some more items should be altered in the survey due to the fact that they can be used more generally than just within IT projects. Putting the word ``Coding'' in brackets, makes the tool useful for non-software development organizations as well. The word ``Working'' should also be added as extra information when the word ``Coding'' is used as a verb.

With the result at hand, we suggested some changes to the items before we collect more data. Table~\ref{fig:lm1} shows the suggested survey for managers and Table~\ref{fig:lm2} shows the suggested survey for developers. Where there is a change made from the Agile Adoption Framework, this is commented at the end of the tables.

\begin{landscape}
\begin{table}
\caption{Suggested Survey for Managers}
\label{fig:lm1}
\includegraphics[scale=0.55]{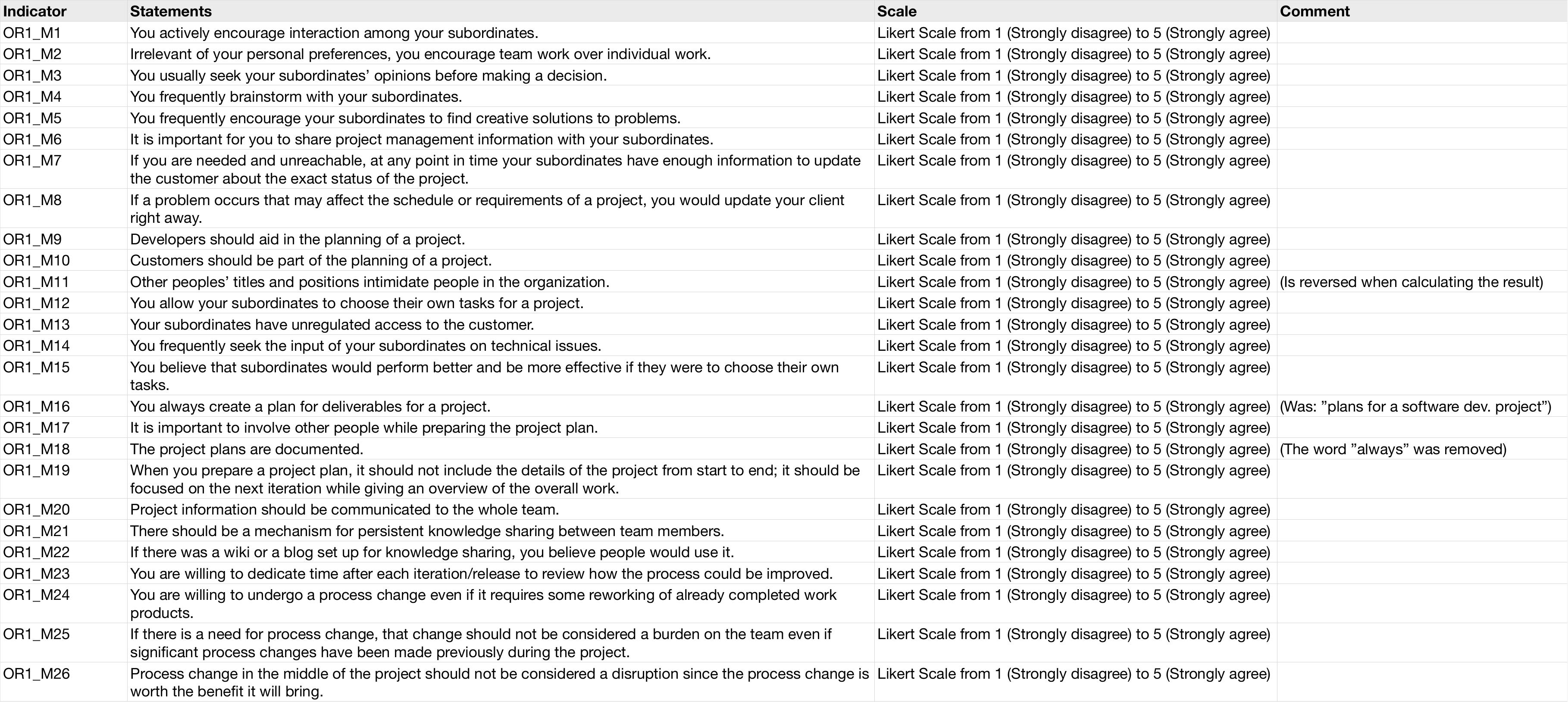}
\end{table}
\end{landscape}
\begin{landscape}
\begin{table}
\caption{Suggested Survey for Developers}
\label{fig:lm2}
\includegraphics[scale=0.5]{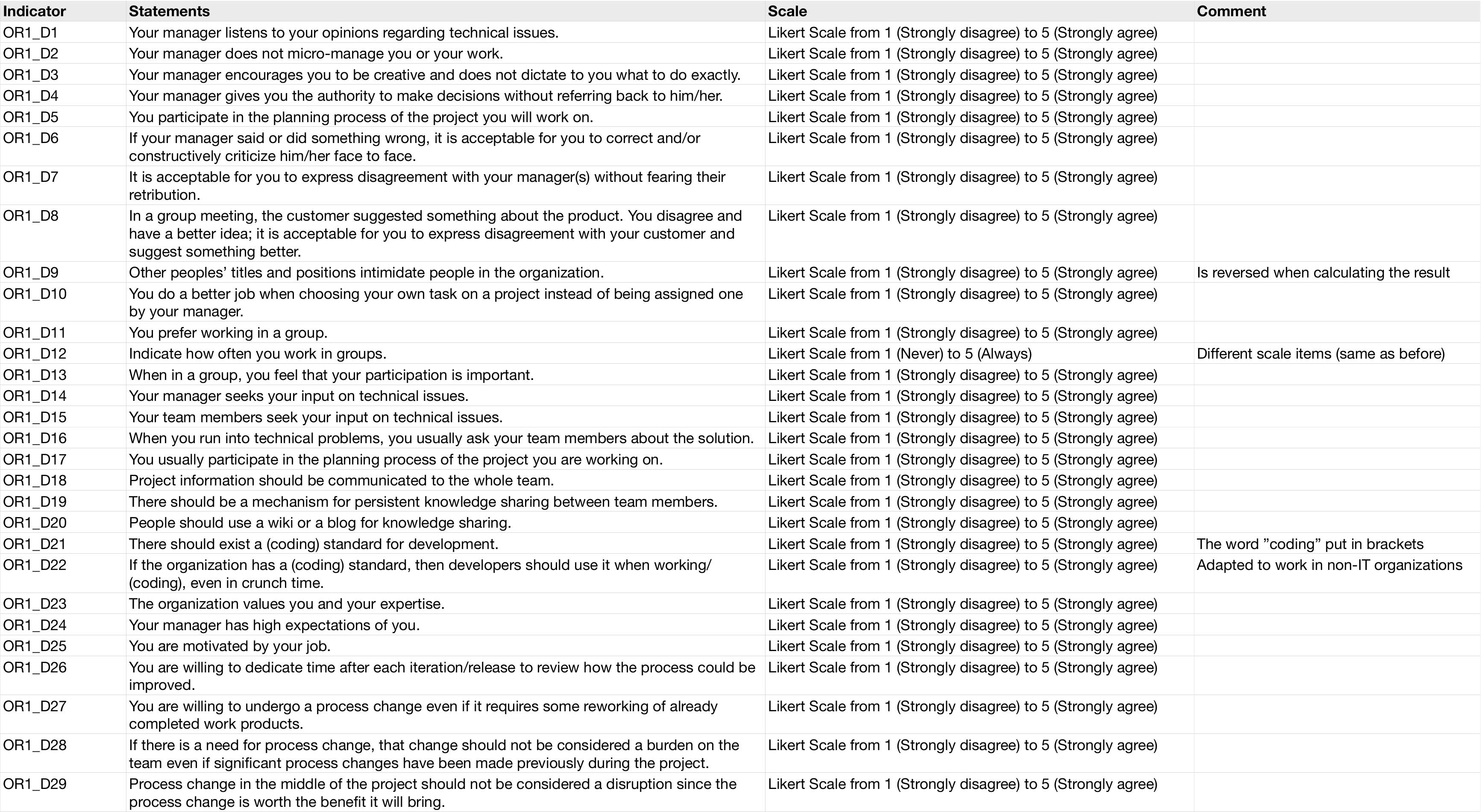}
\end{table}
\end{landscape}

Since we need as much data as possible to run a quantitative statistical analysis, we opted to only use the survey for developers in the exploratory factor analysis, which is the main focus of this study and presented next.

\section{Method}\label{sec:method}

\subsection{Hypothesis Testing}
In this study we want to see if empirical data of the Agile Adoption Framework's Level 1 survey for Developers correspond to Sidky's (\citeyear{sidkyphd}) categorization of agile practices and are reliable and valid according to statistical analyses.

\textbf{Hypothesis:} The Agile Adoption Framework is valid according to quantitative tests for internal consistency and construct validity.

\subsection{Participants}

The sample of the main study consisted of 45 employees from two large multinational US-based companies with 16,000 and 26,000 employees and with revenues of US\$ 4.4 billion and US\$ 13.3 billion respectively. Both stated that they are using agile methods in their participating projects. One of the companies is in the Retail business and the other is in the Consumer Packaged Goods (CPG) industry. However, the groups participating in the research were IT projects within the companies. This study was conducted together with SAP AG\footnote{http://www.sap.com} and they mediated the contacts.

\subsection{Survey}
The survey used in this study was the developer survey presented in the pretest. The survey for developers were put together in an online survey containing 29 items for the team members to answer on a Likert scale from 1 to 5 (where 1 = low agreement to the statement, and 5 =  high agreement). The survey used can be seen in Figure~\ref{fig:lm2}.

\subsection{Procedure}
Two 30 to 45 minute open-ended interviews were conducted with a manager at each company with an overall perspective of their journey towards working agile. The main reason for interviewing managers was to set a psychological contract and get a commitment to making sure the survey were filled in by as many employees as possible, but also, to get the project managers to believe in how the research can help them in the future, and offer to feed the result back to them with recommendations of how to get their group to develop further regarding agility.

The surveys were sent out to the employees via email by their manager. The survey was created as an online survey and the link to it was shared in the email. It was sent to 79 employees and 45 replied, e.g. a response rate of 57\%. This response rate is just above average (55.6\%) within social science research \citep{responserate}. One reminder was sent via email by one of the managers (from one of the organizations). Filling out the survey took approximately 10 minutes and all the questions were compulsory. The actual items can be found in Figure~\ref{fig:lm2}. However, they are named differently but can be found by subtracting 15 from each items in the survey for developers, e.g.\ item Agile41 is item OR1\_D26.

\section{Results}\label{sec:results}
In this section we will present the result of statistical tests for for internal consistency and construct validity. The former will be tested by a Cronbach's $\alpha$ and the latter by exploratory principal factor analysis (or PFA).

However, before these statistical tests we would like to highlight a problem with using the Agile Adoption Framework to measure agility. The terms ``Manager'' and ``Scrum Master/Agile Coach'' could be a source of confusion. Two respondents gave the open-ended feedback of ``We have a PM and an Agile coach. I consider their agile skills to be far apart which lead to some ambiguity when answering questions around `manager'." and ``Some of the questions on my manager are irrelevant or could be misinterpreted. My manager isn't part of the IT organization.'' This ambiguity probably affected the responses since some of the individuals evidently have both a Manager and a Scrum Master.

\subsection{Factor Analysis}

\begin{table}
\caption{Pattern Matrix$^a$ for the agile items.}
\label{fig:patternmatrix}
\centerline{\includegraphics[scale=0.7]{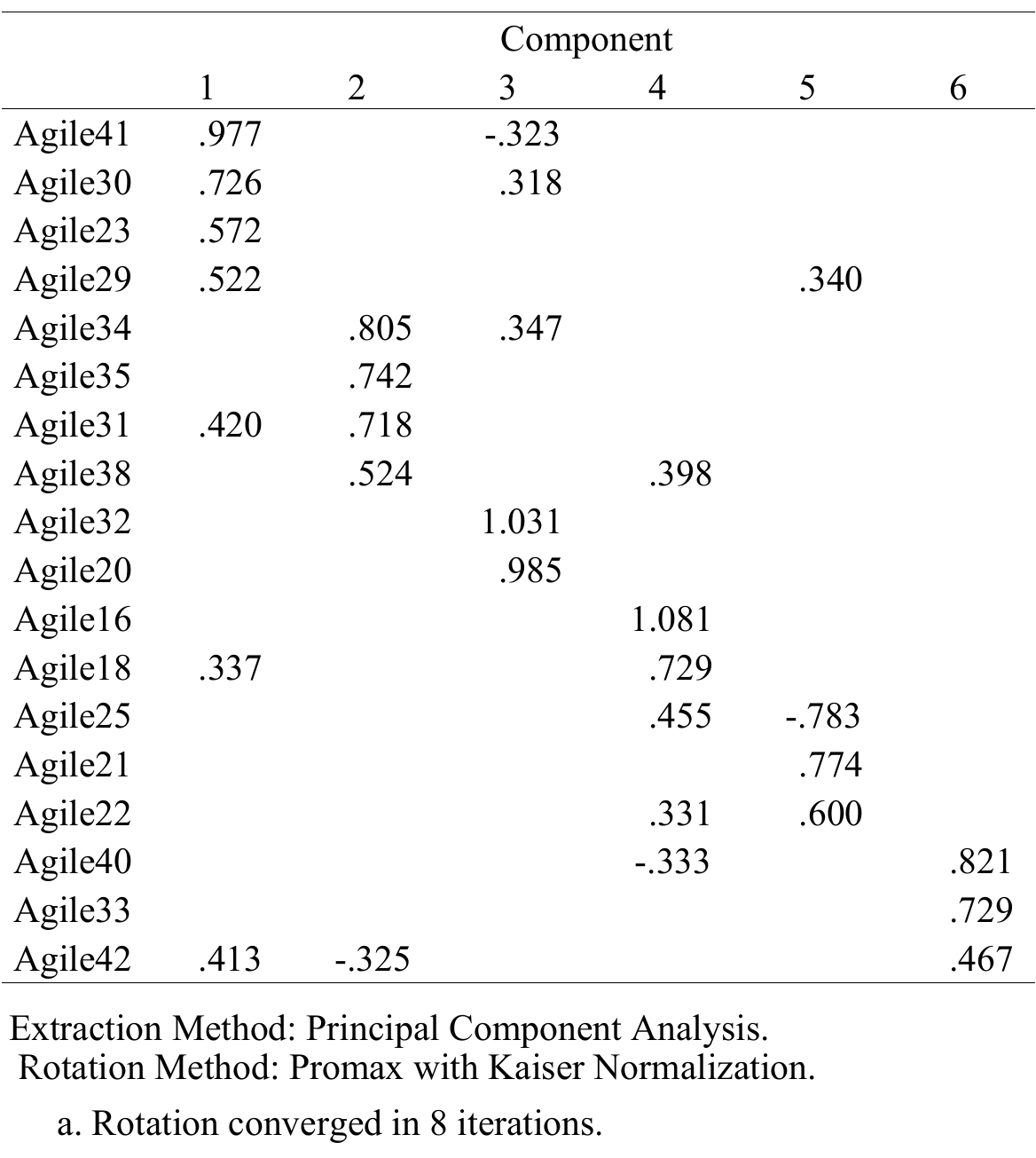}}
\end{table}

\begin{table}
\caption{Structure Matrix for the agile items.}
\label{fig:structurematrix}
\centerline{\includegraphics[scale=0.7]{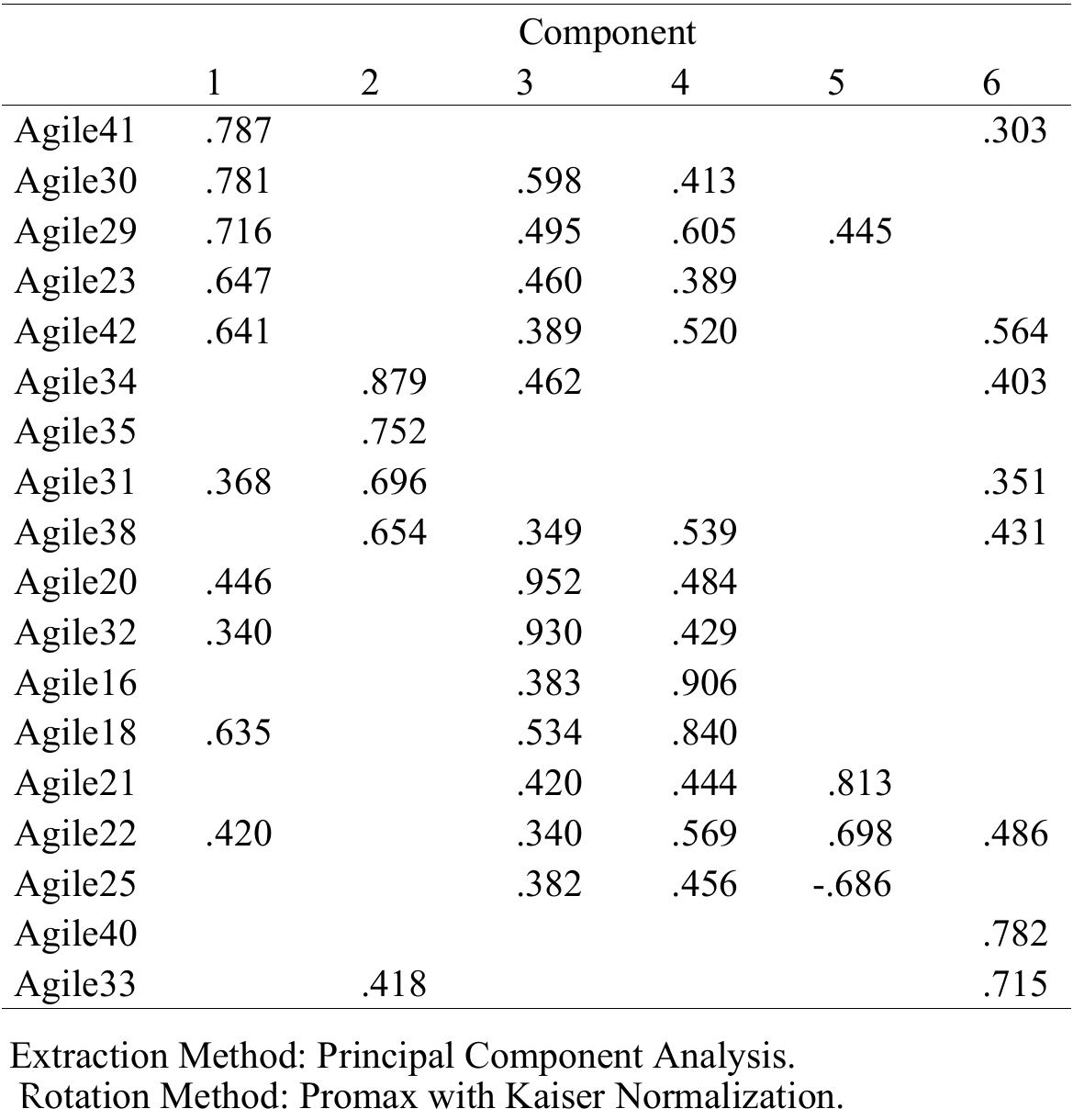}}
\end{table} 

The reason why we used an exploratory principal factor analysis (PFA) instead of a Principal Component Analysis (PCA) is that a PCA is meant to investigate underlying variables in data (i.e. what factors explain most of the variance orthogonally). In a PFA, on the other hand, the variables are grouped if they correlate and explain much of the same variance (i.e. the factors in a scale should not correlate too much or too little if they are considered to explain and measure a construct). A factor analysis is a statistical help to find groups of variables that explain distinct constructs in data. For more details, see e.g.\ \citet{fabrigar}.

The first thing to do when conducting a factor analysis is to make sure the items have the preferences needed for such a method, i.e. they need to be correlated to each other in a way that they can measure the same concept. Testing the Kaiser-Meyer-Olkin Measure of Sampling Adequacy and Bartlett’s Test of Sphericity is a way to do this. The sphericity was significant for the whole set of items, but the Kaiser-Meyer-Olkin Measure of Sampling Adequacy was $<$.5, which implicates removal of items with low correlations to the rest of the items. An Anti-Image table was created and low-value items were removed, i.e. values with Anti-Image correlation $<$.5. After this the Kaiser-Meyer-Olkin Measure of Sampling Adequacy was .713, which is acceptable. The Pattern Matrix is shown in Table~\ref{fig:patternmatrix} and was used to divide the items into new factors. The extraction was based on Eigenvalues $>$1, and the Promax rotation was used since the items might be dependent. As Table~\ref{fig:structurematrix} shows, the items are correlated to more factors than the one with the highest coefficient. This means that the division into factors is not evident and the items do not clearly reflect different factors of agility. However, it should be mentioned that a factor analysis with a sample size of $N = 45$ is generally considered low, but the sample size needed for factor analysis is dependent on e.g. communalities between and over-determination of factors \citep{maccallum}. Communality is the joint variables' possibility to explain variance in a factor. Over-determination of factors is how many factors are included in each variable. In this case, the first factors have a good amount of variables/factor ratio, and factors 3, 4, 5, and 6 include only 2 or 3 variables. The communalities are measured below with a Cronbach's $\alpha$ for each factor. 

\subsubsection{Reliability}
\begin{figure}
\label{fig:resultsfigure}
\centerline{\includegraphics[scale=0.5]{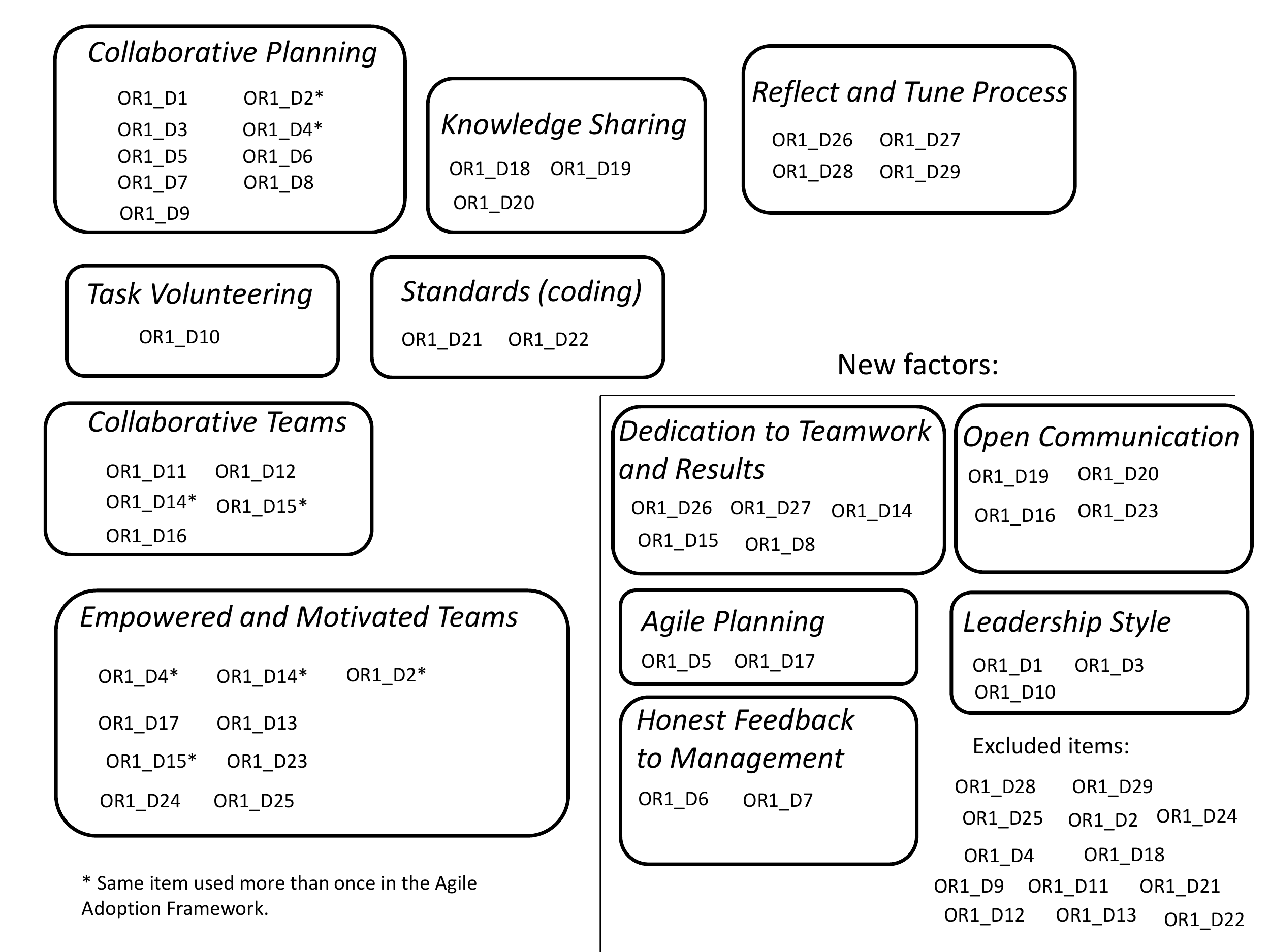}}
\caption{Overview of which items we found support for.}
\end{figure}

After the new factors were created, a Cronbach's $\alpha$ was calculated for each new factor. The factors' $\alpha$ values were: .785, .761, .925, .707, .773, and .470 respectively. Values between .7 and .8 are acceptable for surveys and below .5 is unacceptable since the questions then do not cover the same construct they set out to investigate \citep{cronbach}. The last factor (Factor 6) was therefore removed from the rest of the analysis. The other five factors were divided and named as follows: ``Dedication to Teamwork and Results'' (Agile41, Agile42, Agile30, Agile23 and Agile29), ``Open Communication'' (Agile34, Agile35, Agile31 and Agile38), ``Agile Planning'' (Agile32 and Agile20), ``Leadership Style'' (Agile16, Agile18 and Agile25), and ``Honest Feedback to Management'' (Agile21 and Agile22). Figure~\ref{fig:resultsfigure} shows an overview of the items we found support for. 

Since it was not possible to verify the agile practices division made by \citet{sidkyphd} by conducting a factor analysis on data, \emph{the hypothesis was rejected}. 

\section{Discussion}\label{sec:discussion}
In this study we first tested how practitioners rate the use of the Agile Adoption Framework through a focus group. The result of this was positive. However, the statistical tests did not support the categorization of factors in the framework and can therefore not be considered to measure distinct constructs (i.e.\ being a valid measurement for agility, in this case).

The pretest showed that the teams found the categories of the Agile Adoption Framework relevant and measured how the teams worked in their new process. However, the statistical analyses suggest this measurement needs more work in order to be a valid measurement of agile practices implemented in a team. This can be due to a diversity of reasons; first, a cultural change in an organization is by definition hard to assess and very contextual. Perhaps this set of items do not reflect what agility is, however, we believe a set of items that considers a cultural as well as a behavioral dimension could be constructed in the future. 

Even if the Agile Adoption Framework does not measure the agility construct as expected and therefore the hypothesis was rejected, the items were still developed and checked for content validity by \citet{sidkyphd}, i.e. it is coherent with what some practitioners define as ``agility''. However, as mentioned in the introduction, a statistical analysis must support the items to be considered a valid measurement. None of the categories defined in the Agile Adoption Framework were statistically verified. Even though this was the case, the set of items that Sidky generated are covering much of the behavior connected to agile development processes. Practitioners seem to be keen on measuring agility since they want to show proof of their success for a set of reasons, however, this does not mean the measurements really reflect agility as shown by this study. 

Another possible explanation could be that our sample is too small (or skewed) to say that Sidky's categories are not supported. However, when constructing a survey tool (or ``scale'' in psychology) one must verify the categorizations made qualitatively through a quantitative validation. Hence, any of the mentioned agile maturity models need more development before they can be considered reliable. Furthermore, to trust the result in this study another independent PFA should be done and compared to this one. If two or more independent PFAs give the same result, we would be certain our results hold. Therefore, this result is only a first step in creating a validated tool. 

Over the last decade, a diversity of agile maturity models have surfaced, as described in the introduction \citep{lepp}. It is a pity that researchers keep inventing new ones instead of validating (or even merging) existing tools to actually find a couple that works. Even the same year as the work of \citet{lepp} was presented, more models have been suggested (by e.g.\ \citet{soundphd}). New ideas and models are good but in this context what is really needed is to validate the existing ones so practitioners can be comfortable using them. 

However, there is another fundamental issue with agile maturity models. Even if we can develop a statistically valid set of items to measure agile practices, a team's score on such a scale might not reflect what is actually meant by an agile team. The term ``agile process'' is undefined and many researchers and practitioners have their own definition and perception of what it exactly means. It is clear, though, that agile processes are not such a set of hands-on practices. Since agile principles are more about culture than a set of implemented methods, maybe a maturity level approach is not the way to go. Or we need another focus in the measurements that include cultural assessments instead of degree of used practices. 

The fact that the different agile maturity models have the same agile practice in a range of different levels \citep{lepp}, also indicates that the maturity levels of agility are not evident. Maybe this is a syndrome of not letting go of the control mechanisms that agile principles suggest should be more in the periphery. Since agile methods are more about people and culture we suggest social psychological measurements are more appropriate if organizations want to measure their level of agility. The only study we found on social psychology and agile development processes is the article \emph{Perceptive Agile Measurement:
New Instruments for Quantitative Studies in the Pursuit of the Social-Psychological Effect of Agile Practices} by \citet{so}. Their work deserves more attention since they created a tool and validated it on a sample of $N=227$. Since we want to measure agility in organizations, this tool will make such a measurement feasible since it excludes specific practices and focuses on behavior connected to the underlying agile principles. 

The Agile Adoption Framework is intended to assess agility before these ideas have been introduced into the organization, however, we believe an organization that has no clue what the wording ``Agile Processes'' means could still be agile in their ways of working. We also believe the opposite is true; an organization can have implemented agile practices without really being agile. Therefore, the measurement of agility should not be dependent on what the organization calls a ``manager'', ``team lead'' or ``agile coach'' etc., but focus on what these people are doing. This is a threat to this study since questions regarding the manager were reported to be hard to interpret. However, this is also part of our critique we just mentioned regarding building a tool that is not dependent on such jargon. The other aspects of the tool did not form factors anyways, but we have suggested new categories for the Agile Adoption Framework. These were:  ``Dedication to Teamwork and Results'', ``Open Communication'', ``Agile Planning'', ``Leadership Style'', and ``Honest Feedback to Management''. This makes the Agile Adoption Framework \citep{sidkyphd} one of few agile maturity level now partially statistically validated (on level 1 in one of the step described by Sidky). However, the questions still includes some ambiguity regarding manager and agile leader. Furthermore, the Agile Adoption Framework uses the same items to assess both results for developers and managers, which makes statistical analysis cumbersome. However, as mentioned, in our validation we also only used the survey for developers.

Sidky's tool was not intended to measure agility of a team but agile potential. This separation of perspectives is the reason why his survey for managers does not include Agile Management concepts like the definition of ``Done''. We argue, though, that a team can be agile without having implemented agile practices and therefore this type of Boolean response to if a team is agile or not before the measurement is conducted, does not cover what agility is, according to us.  

We should also mention that the largest contribution by \citet{sidkyphd}, as we see it, is not his agile team level potential assessment, but the overall items regarding a go\slash no go decision process at an early stage to see if agile methods is a good idea for a specific organization. This part is not presented in this study but is a great contribution to the field. 

We believe the work of \citet{so} could be combined with the Agile Adoption Framework to reflect more aspects of agility in such an assessment. Then the dimensions presented in the Perceptive Agile Measurement:

\begin{itemize}
\item Iteration Planning
\item Iterative Development
\item Continuous Integration \& Testing
\item Stand-Up Meetings
\item Customer Access
\item Customer Acceptance Tests
\item Retrospectives
\item Collocation
\end{itemize}

can be assessed jointly with the output of this study:

\begin{itemize}
\item Dedication to Teamwork and Results
\item Open Communication
\item Agile Planning
\item Leadership Style
\item Honest Feedback to Management
\end{itemize}

which we believe create a powerful and useful tool that can give teams focus points to improve. However, we believe more dimensions are still needed and can be taken from other management fields. One of these aspects that certainly affect agile adoption is, for example, to measure innovation propensity \citep{dobni}. However, to measure all aspects of an organization in relation to agility will take time and there is always a tradeoff between doing these time-consuming expert assessment (like Sidky's entire tool) or only measuring a subset to obtain indications of focus areas, like suggested in this study.

\subsection{Validity Threats}\label{sec:val}
Our result and therefore also our conclusions could be due to the fact that our sample is too small or that Sidky's (\citeyear{sidkyphd}) tool is not possible to use as a quantitative tool. The ambiguity of the different perspectives (where Sidky wants to measure agile potential and we aim to measure current agility) is also a threat to validity. We have also questioned the usefulness of using these types of agile maturity models since they do not take culture, or the purpose of using agile methods, into account. Furthermore, we have used a Principal Factor Analysis in this study which is used under the assumption that the observed variables are a linear combination of the factors. While doing this we also assume that a Likert scale generates interval data. These aspects are, however, more a part of a general discussion on the usefulness of some statistical models in social science.

\section{Conclusions and Future Work}\label{sec:future}
In conclusion, this study has shown that quantitative data do not support the categorization of a subset of items in the Agile Adoption Framework. 
This is not surprising since no quantitative validation has been done on the tool, but troublesome for both research and practice since no thoroughly validated tool is a their disposal. Since this is the case researchers cannot correlate quantitative agile maturity measurements to other variables in Software Engineering research and be confident that the results are correct. Practitioners cannot either use these tools to guide their journey towards agility. In order to create a validated survey, the items must be iterated with real data until supported and reliable. By first doing a pretest with a small sample ($N=15$) we qualitatively validated the items. After a few alterations we ran a factor analysis and a reliability test on the tool ($N=45$). Data did not support the division of a subset of items selected from the Agile Adoption Framework. However, the data gave new categorizations of the items in the Agile Adoption Framework. As far as we know, this gives one of the first partially validated agile maturity model.

To summarize this study has contributed with:

\begin{enumerate}
\item A positive result/feedback from practitioners on the usage if the Agile Adoption Framework as measure of current agility (instead of agile potential), in a pretest case study.
\item Evolvement of the method of the Agile Adoption Framework to include a Likert scale evaluation survey filled out by all the team members and not just by the assessor/researcher and connect confidence intervals to the item results. This way of assessing agility is less time consuming for the assessor.
\item Validation tests for internal consistency and construct validity on the Agile Adoption Framework on additional data suggest the data collected did not support the way the indicators are related to the agile practices (on Level 1) in the framework under investigation.
\item This study finds support for a new division of items to measure agility but concludes that much validation is needed to even state that the items measure the agile practices. Furthermore, we question agile maturity models as a good way to assess agility and propose that tools look more into other dimensions like culture and innovation propensity. 
\item This study also highlights the tradeoff between quick quantitative measurements to guide agile adoption that is much wanted by practitioners and time-consuming contextual and more qualitative assessments in organizations that might be closer to the real situation. 
\end{enumerate}

We believe the next step for this kind of research would be to combine the items from many agile maturity models and see where they overlap. These items should then be subjected to the same analysis conducted in this study with a larger data set. Obviously, the larger the sample the better when validating a tool and it would be good to validate all maturity models (including the Agile Adoption Framework) with an even larger sample. However, we believe new separate agile maturity models have ceased to contribute to the development of measuring agility, and we want to stress the importance of creating one validated combination instead. We also see the importance of adding other dimensions than agile practices to these measurements, such as validated measurements of organizational culture and innovation propensity.

\section*{Acknowledgements}
This study was conducted jointly with SAP AG\footnote{http://www.sap.com}, and we would especially like to thank Jan Musil at SAP America Inc. We would also like to thank the SAP customers who were willing to share information. Volvo Logistics, Pasi Moisander, Karin Scholes, and Kristin Boissonneau Gren (without your goodwill this work could not have been done).

\bibliographystyle{model5-names}
\bibliography{references}
\newpage

\end{document}